\documentclass[journal=jctcce,manuscript=article]{achemso}
\usepackage[english]{babel}
\usepackage[utf8]{inputenc}
\usepackage[T1]{fontenc}
\usepackage{graphicx}
\usepackage{amsmath, amsthm, amssymb, amsfonts}
\usepackage{color}

\title{Zero-point energy leakage in Quantum Thermal Bath molecular dynamics simulations}

\author{Fabien Brieuc}
\affiliation{Laboratoire Structures Propriétés et Modélisation des Solides, CentraleSupélec, CNRS, Université Paris-Saclay,  92295 Châtenay-Malabry, France}
\altaffiliation{These authors contributed equally to this work}
\author{Yael Bronstein}
\affiliation{Sorbonne Universités, UPMC Université Paris 06, CNRS, Institut des Nanosciences de Paris, 4 Place Jussieu 75252 Paris, France}
\altaffiliation{These authors contributed equally to this work}
\author{Hichem Dammak}
\affiliation{Laboratoire Structures Propriétés et Modélisation des Solides, CentraleSupélec, CNRS, Université Paris-Saclay,  92295 Châtenay-Malabry, France}
\alsoaffiliation{Laboratoire des Solides Irradiés, École Polytechnique, CNRS, CEA, Université Paris-Saclay, 91128 Palaiseau, France}
\email{hichem.dammak@centralesupelec.fr}
\author{Philippe Depondt}
\affiliation{Sorbonne Universités, UPMC Université Paris 06, CNRS, Institut des Nanosciences de Paris, 4 Place Jussieu 75252 Paris, France}
\author{Fabio Finocchi}
\email{fabio.finocchi@insp.upmc.fr}
\affiliation{Sorbonne Universités, UPMC Université Paris 06, CNRS, Institut des Nanosciences de Paris, 4 Place Jussieu 75252 Paris, France}
\author{Marc Hayoun}
\affiliation{Laboratoire des Solides Irradiés, École Polytechnique, CNRS, CEA, Université Paris-Saclay, 91128 Palaiseau, France}

\begin{document}
\begin{abstract}
The quantum thermal bath (QTB) has been presented as an alternative to path-integral based methods to introduce nuclear quantum effects in molecular dynamics simulations. The method has proved to be efficient, yielding accurate results for various systems. However, the QTB method is prone to zero-point energy leakage (ZPEL) in highly anharmonic systems. This is a well known problem in methods based on classical trajectories where part of the energy of the high frequency modes is transferred to the low frequency modes leading to a wrong energy distribution. In some cases, the ZPEL can have dramatic consequences on the properties of the system. Thus, we investigate the ZPEL by testing the QTB method on selected systems with increasing complexity in order to study the conditions and the parameters that influence the leakage. We also analyze the consequences of the ZPEL on the structural and vibrational properties of the system. We find that the leakage is particularly dependent on the damping coefficient and that increasing its value can reduce, and in some cases, completely remove the ZPEL. When using sufficiently high values for the damping coefficient, the expected energy distribution among the vibrational modes is ensured. In this case, the QTB method gives very encouraging results. In particular, the structural properties are well reproduced. The dynamical properties should be regarded with caution although valuable information can still be extracted from the vibrational spectrum, even for large values of the damping term.
\end{abstract}

%************** Introduction *****************
\section{Introduction} \label{sec_intro}
The molecular dynamics (MD) simulation technique is a powerful tool to investigate the properties of complex atomic systems. 
At low temperature and/or in systems containing light elements such as hydrogen, nuclear quantum effects can play a major role on the behavior of the system. However, it is currently a computational challenge to account for the quantum nature of nuclei in MD simulations.

Over the past years, several techniques have been proposed to deal with this issue. Among them, the most common methods are based on the path integral formalism such as path integral molecular dynamics (PIMD). In this formalism, each quantum nucleus is described by a ring of classical monomers (or beads) connected through harmonic springs \cite{ceperley1995,marx1999,tuckerman2010}. When the number of beads is large enough, the statistical averages converge towards the exact quantum result. However, in order to compute time correlation functions, approximate methods such as centroid MD\cite{cmd} or ring-polymer MD\cite{craig2004,braams2006} are needed. These path-integral methods are computationally demanding when the number of beads increases, in particular at low temperature.

Recently, alternative methods based on a modified Langevin equation have been proposed \cite{dammak2009,ceriotti2009}. Among them, the quantum thermal bath (QTB)\cite{dammak2009} is an approximate yet efficient method to include nuclear quantum effects in MD simulations. Although exact only in the case of a system of harmonic oscillators, the QTB provides satisfactory results in many anharmonic systems.\cite{calvo2012,dammak2012,savin2012,basire2013,bronstein2014,bronstein2016} A first advantage is its implementation without any additional computational cost compared to standard MD. Hence, large and complex systems can be in principle treated by QTB-MD. Moreover, the method can give information about dynamical properties of the system. Finally, its formulation is not system-dependent, in particular, no knowledge of the system's vibrational density of states is needed beforehand. However, the QTB method has several drawbacks. First, the method can fail when dealing with highly anharmonic systems.\cite{dammak2011} Second, the QTB technique is prone to zero-point energy leakage (ZPEL), like any other method based on classical trajectories.\cite{bennun1994} 

The ZPEL is a known problem where a part of the energy of the high-frequency modes is transferred to the low-frequency ones which is due to the classical nature of MD trajectories. The ZPEL was observed in many different systems (water clusters and liquid water, Lennard-Jones systems, ...)\cite{habershon2009,ceriotti2010,ganeshan2013,bedoya2014,hernandez2015}; in particular, in the case of the QTB, the ZPEL has been recently pointed out by Bedoya-Martinez and coworkers.\cite{bedoya2014} However, no systematic or general study of ZPEL within the QTB framework has been done up to now.

Several solutions to the ZPEL problem within QTB-MD simulations have recently been suggested. Bedoya-Martinez and coworkers tried to modify the noise power spectrum in order to obtain the expected energy distribution. However, this solution is system-dependent and only worked for weakly anharmonic systems.\cite{bedoya2014} Ganeshan and coworkers proposed a deterministic approach to suppress ZPEL, which unfortunately requires the knowledge of the vibration normal coordinates prior to the simulation.\cite{ganeshan2013}

Here, we investigate the conditions leading to the ZPEL within QTB-MD simulations in various systems in order to get a better understanding of the validity of the QTB method. More precisely, we focus on the conditions and the parameters that influence the ZPEL and on the consequences for the system's properties. After a brief presentation of the QTB method, we study selected anharmonic systems with increasing complexity. First, we investigate two simple models: two coupled harmonic oscillators and a one-dimensional chain of atoms. Then, we focus on more realistic systems, a Lennard-Jones aluminium crystal and the phase transitions in BaTiO$_3$. In the last section, we discuss our results and their implications.

%********** Section1 : The QTB method  ****************
\section{The Quantum Thermal Bath method} \label{sec_1}
{The QTB method is based on a modification of the Langevin thermostat in order to include nuclear quantum effects in MD simulations.}
Both in the standard (i.e. classical) Langevin thermostat and in the QTB method, the equation of motion for one degree of freedom $x$ of mass $m$ and submitted to the {internal} force $f(x)$ reads:\cite{dammak2009}
\begin{equation}
m\ddot{x}= f(x) - m\gamma\, \dot{x}+ R(t)     \label{eq-EOM}
\end{equation}
The last two terms correspond to the friction and stochastic forces of the thermostat, respectively.

{
The random force is described by a stationary stochastic process $R(t)$ whose distribution is Gaussian with zero mean:
\begin{align}
<R(t)>&=0\\
<R(t)R(t+\tau)>&=\int_{-\infty}^{+\infty}I_R(\omega,T)\, \text{e}^{-i\omega\tau}\, \frac{\text{d}\omega}{2\pi}.
\label{eq:WK_th}
\end{align}
Equation \ref{eq:WK_th} is the Wiener-Khinchin theorem, which relates the autocorrelation function $<R(t)R(t+\tau)>$ of the stochastic process to its power spectral density (PSD) $I_R(\omega,T)$ at temperature $T$. The dynamical properties obtained using eq. \ref{eq-EOM} are directly related to this PSD. 
The closely related PSD of the position, $I_x$, is obtained from the fluctuation-dissipation theorem \cite{Kubo}, 
which reads in the classical case:
\begin{equation}
\tilde{\chi}^{''}(\omega) = \frac{\omega}{2k_BT} I_{x}(\omega,T)
\end{equation}
with $\tilde{\chi}^{''}(\omega)$ the imaginary part of the susceptibility $\tilde{\chi}(\omega)$ that connects the Fourier transform of the position $\tilde{x}(\omega)$ to the Fourier transform of the random force $\tilde{R}(\omega)$ within the linear response theory: 
\begin{equation}
\tilde{x}(\omega)=\tilde{\chi}(\omega)\tilde{R}(\omega)  \label{suc}
\end{equation}
From this expression, we obtain a linear relation between the PSD of the position, $I_x$, and the PSD of the stochastic force, $I_R$:
\begin{equation}
I_x(\omega,T) = |\tilde{\chi}(\omega)|^2 I_R(\omega,T)
\end{equation}
and the fluctuation-dissipation theorem can be rewritten as follows:
\begin{equation}
I_R(\omega,T)=\frac{2k_BT}{\omega}\frac{\tilde{\chi}^{''}(\omega)}{|\tilde{\chi}(\omega)|^2}. \label{eq-FDT}
\end{equation}
In the case of an harmonic oscillator with an angular frequency $\omega_0$, 
using eq. \ref{eq-EOM} and \ref{suc} the susceptibility writes
\begin{equation}
\tilde{\chi}(\omega)=\frac{1}{m\left[\omega_0^2-\omega^2+i\gamma\omega\right]}.
\end{equation}
By introducing this expression in eq. \ref{eq-FDT}, the PSD of the random force is obtained in the classical case
as a white noise:
\begin{equation} \label{eq:cl-PSD}
I_{R}(\omega,T) = 2m\gamma \,k_BT  \hspace*{1cm} \forall \omega .
\end{equation} 
By using this expression and the Wiener-Khinchin theorem (eq. \ref{eq:WK_th}) in eq. \ref{eq-EOM}, the standard Langevin dynamics is obtained. In this case, the equipartition of the energy is ensured, and all harmonic vibrational modes have the same average energy ($k_BT$), which is independent of the angular frequency $\omega$.

In the quantum case, the average energy of a vibrational mode is given by
\begin{equation}
\theta(\omega,T)=\hbar \omega \left[\frac{1}{2} + \frac{1}{\exp\left(\frac{\hbar\omega}{k_BT}\right)-1}\right] 
\label{eq-theta}
\end{equation} 
in the harmonic approximation. The main idea of the QTB method is to replace the PSD of the classical random force (\ref{eq:cl-PSD}) by the one corresponding to the energy distribution of eq. \ref{eq-theta}. 
This is done, in practice, by using the quantum version of the fluctuation-dissipation 
theorem as developed by Callen and Welton \cite{CallenWelton} and reviewed by Kubo \cite{Kubo}, which gives:
\begin{equation}
\tilde{\chi}^{''}(\omega) = \frac{\omega}{2\theta(\omega,T)} I_{x}(\omega,T)
\end{equation} 
and, through the Wiener-Khinchin theorem, leads to the PSD of the colored noise $R(t)$ as used in the QTB method:
\begin{equation}
I_{R}(\omega,T) = 2m\gamma\, \theta(\omega,T) \label{eq-IR}
\end{equation}
In contrast to the Langevin thermostat, $I_{R}$ is $\omega$ dependent and the random force $R(t)$ is obtained using the procedure \cite{maradudin,chalopin} described in Appendix \ref{appendixA}.
}

The use of an angular frequency cut-off $\omega_{\text{cut}}$ is necessary during the generation of these random forces \cite{BarratRodney} because the average energy of a harmonic oscillator diverges at high frequencies. Thus, the QTB method contains two free parameters : the friction coefficient $\gamma$ and the angular frequency cut-off $\omega_{\text{cut}}$. The values of these parameters must be carefully chosen. 
When using the Langevin thermostat, it is generally assumed that the friction coefficient $\gamma$ 
has to be small enough so that the forces associated with the thermostat 
do not significantly perturb the dynamics of the system.\cite{Pastorino2007}
Moreover, as already stated by Barrat and Rodney \cite{BarratRodney}, $\omega_{\text{cut}}$ must be chosen of the order of a few times the highest angular frequency observed in the system to prevent the inclusion of non-physical high frequency modes. Too high values for $\omega_{\text{cut}}$ and $\gamma$ could lead to the divergence of the total energy.\cite{BarratRodney}
In the simulations that are presented in this work, we found that a reasonable value for $\omega_\text{cut}$ is approximately $2\omega_{\text{max}}$ with $\omega_{\text{max}}$ being the highest angular frequency in the system.
In the following, we focus on the results of QTB-MD simulations when increasing the friction coefficient $\gamma$ 
in eq.\ref{eq-EOM} and, consistently, in the power spectrum of the stochastic force (eqs. \ref{eq:cl-PSD} and \ref{eq-IR}).

%************ Section2 : 2 coupled oscillators  *****************
\section{Model systems} \label{sec_2}
\subsection{Coupled harmonic oscillators}
\begin{figure}[t]
\includegraphics[width=3.33in]{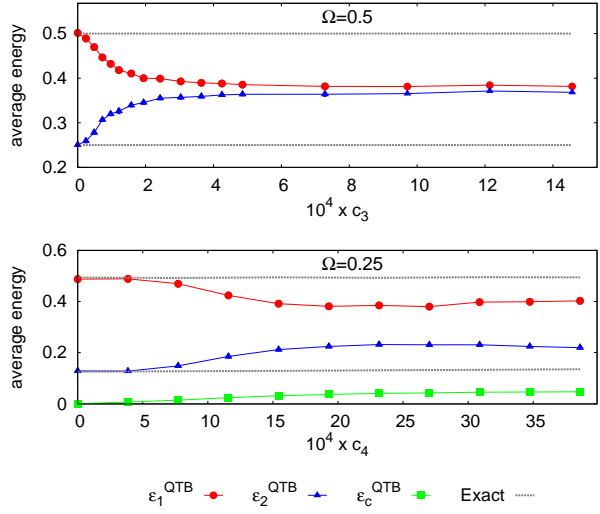}
\caption{Average energies, $\epsilon_1$ and $\epsilon_2$, of the two oscillators,  and average coupling energy $\epsilon_c$ computed by QTB-MD as a function of the intensity of the coupling constants $c_3$ and $c_4$. 
Top panel: cubic coupling ($c_3 \neq 0$, $c_4=0$) with $\Omega = 0.5$. 
Lower panel: quartic coupling ($c_4 \neq 0$, $c_3=0$) with $\Omega = 0.25$. 
By symmetry, $\epsilon_c  = 0$ in the cubic case. }
%The exact quantum results (grey dashed line) are obtained by numerically solving the stationary Schr\"odinger equation. }
\label{fig:sec2:ener}
\end{figure}

In this section, we study the behaviour of QTB-MD on a simple model consisting of two coupled one-dimensional harmonic oscillators. Thanks to the small number of degrees of freedom, we can directly compare the QTB-MD results with the numerical solution of the time-independent Schr\"odinger equation here.
The system is described by the Hamiltonian $H$:
\begin{equation}
H=\frac{1}{2}m\dot{x}_1^2 + \frac{1}{2}m\omega_1^2 x_1^2 + \frac{1}{2}m\dot{x}_2^2 + \frac{1}{2}m\omega_2^2 x_2^2 +  C_3(x_1 - x_2)^3 + C_4(x_1-x_2)^4 
\end{equation}
where $x_1$ and $x_2$ are the positions of the two oscillators, $\omega_1$ and $\omega_2$ are their angular frequencies, $m$ is their mass and $C_3$ and $C_4$ are coupling constants.
The Hamiltonian $H$ can be written in a dimensionless form, $\tilde{H}=H/\hbar\omega_1$, so that:
\begin{equation}
\tilde{H} = \frac{\dot{q}_1^2}{2} + \frac{q_1^2}{2} + \frac{\dot{q}_2^2}{2} + \Omega^2\frac{q_2^2}{2}  + c_3(q_1-q_2)^3 + c_4(q_1 - q_2)^4
\end{equation}
where the following variables are used:
\begin{equation}
\Omega = \frac{\omega_2}{\omega_1}, \quad \xi = \sqrt{\frac{\hbar}{m\omega_1}}, \quad q_i = \frac{x_i}{\xi}, \quad  c_3= \frac{C_3\xi^3}{\hbar\omega_1} , \quad c_4 	= \frac{C_4\xi^4}{\hbar\omega_1}, \quad t^* = \omega_1 t,\quad \dot{q}_i = \frac{\mathrm{d}q_i}{\mathrm{d}t^*}.
\label{eq:sec2:varchange}
\end{equation}
$q_1$ and $q_2$ are the reduced positions of the two oscillators and $\Omega$ is the ratio of the frequencies of the two oscillators (we set $\omega_1>\omega_2$). The non-linear coupling terms introduce a controllable degree of anharmonicity in the system which in turn leads to a clear illustration of the ZPEL within the QTB method and allows for the analysis of the conditions leading to this phenomenon. In particular, we study here the influence of cubic and quartic coupling terms on the energies of the oscillators $\epsilon_1$ and $\epsilon_2$, that are:
\begin{equation}
\epsilon_1=\left\langle \frac{\dot{q}_1^2}{2}\right\rangle+\left\langle \frac{q_1^2}{2}\right\rangle, \quad \epsilon_2=\left\langle \frac{\dot{q}_2^2}{2} \right\rangle +\Omega^2\left\langle \frac{q_2^2}{2}\right\rangle.
\label{energies}
\end{equation}
The QTB-MD simulations were performed with a friction coefficient $\gamma = 4\times 10^{-4}\,\omega_1$, a cut-off frequency $\omega_{\text{cut}} = 2\omega_1$ and a time step $\delta t = 0.05\omega_1^{-1}$. Average values are computed using at least 30 independent trajectories that are $10^7$ time steps long each. The ratio $\Omega$ is varied in the 0.05--0.8 range and the parameters $c_3$ and $c_4$ are varied in the 0--$25\times 10^{-4}$ and 0--$40\times 10^{-4}$ ranges respectively, so that we cover a large range of coupling energies (figure \ref{fig:sec2:ener}).  
The temperature is set to $k_B T = 0.03\, \hbar \omega_1$ (e.g. $T \sim 60$ K if $\omega_1 = 2\pi \times 40$ THz) so that the thermal energy contribution to the energies of the oscillators is negligible with respect to their zero-point energies. 

%-------------------------
% FIGURE 1
%-------------------------

The exact quantum calculation shows that the energies of the oscillators are almost independent of the anharmonic coupling intensities for the range of coupling values studied here and are equal to their zero-point energies; hence, in reduced units, $\epsilon_1 = 0.5$ and $\epsilon_2 = \Omega/2$.
Figure \ref{fig:sec2:ener} shows the average energies obtained with the QTB method in two distinct cases: $\Omega=0.5$ with only a cubic coupling and $\Omega=0.25$ with only a quartic coupling.
As expected, in the uncoupled case, i.e. $c_3=0$ and $c_4=0$, the QTB method gives the expected quantum energies for the two oscillators, corresponding to their zero-point energies. 
In contrast, when the coupling constants $c_3$ or $c_4$ are increased, the QTB-MD energies diverge from the exact results: part of the energy of oscillator 1 is transferred into oscillator 2, hence, ZPEL occurs.
In the following, we investigate how the ZPEL depends on the three parameters ($\Omega$,$c_3$,$c_4$) that define the Hamiltonian. 

%-------------------------
% FIGURE 2
%-------------------------
\begin{figure}[t]
\includegraphics[width=3.33in]{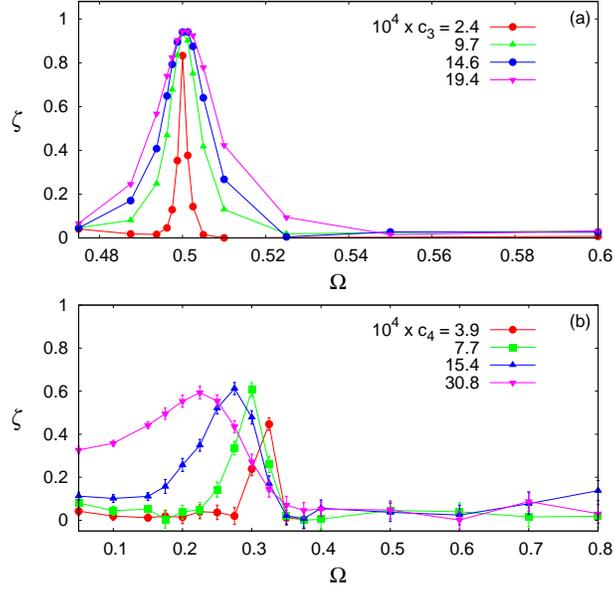}
\caption{Zero-point energy leakage quantified through the deviation factor $\zeta$ (eq. \ref{eq:zeta}) as a function of the ratio of frequencies $\Omega$ (eq. \eqref{eq:sec2:varchange}). Top panel: cubic coupling ($c_3\neq 0$, $c_4=0$). Lower panel: quartic coupling ($c_4 \neq 0$, $c_3=0$).}
\label{fig:sec2:zpel_vs_coupling}
\end{figure}

First, we adopt the following deviation factor $\zeta$ in order to quantify the ZPEL: 
\begin{equation}
\zeta = \frac{\Delta\epsilon^\text{exact}-\Delta\epsilon^\text{QTB}}{\Delta\epsilon^\text{exact}}=\frac{(\epsilon_1^\text{exact} - \epsilon_2^\text{exact}) - (\epsilon_1^\text{QTB}-\epsilon_2^\text{QTB})}{\epsilon_1^\text{exact}-\epsilon_2^\text{exact}}
\label{eq:zeta}
\end{equation}
%
%-----------------------------
% FIGURE 3
%----------------------------
\begin{figure}[t]
\centering
\includegraphics[width=4.4in]{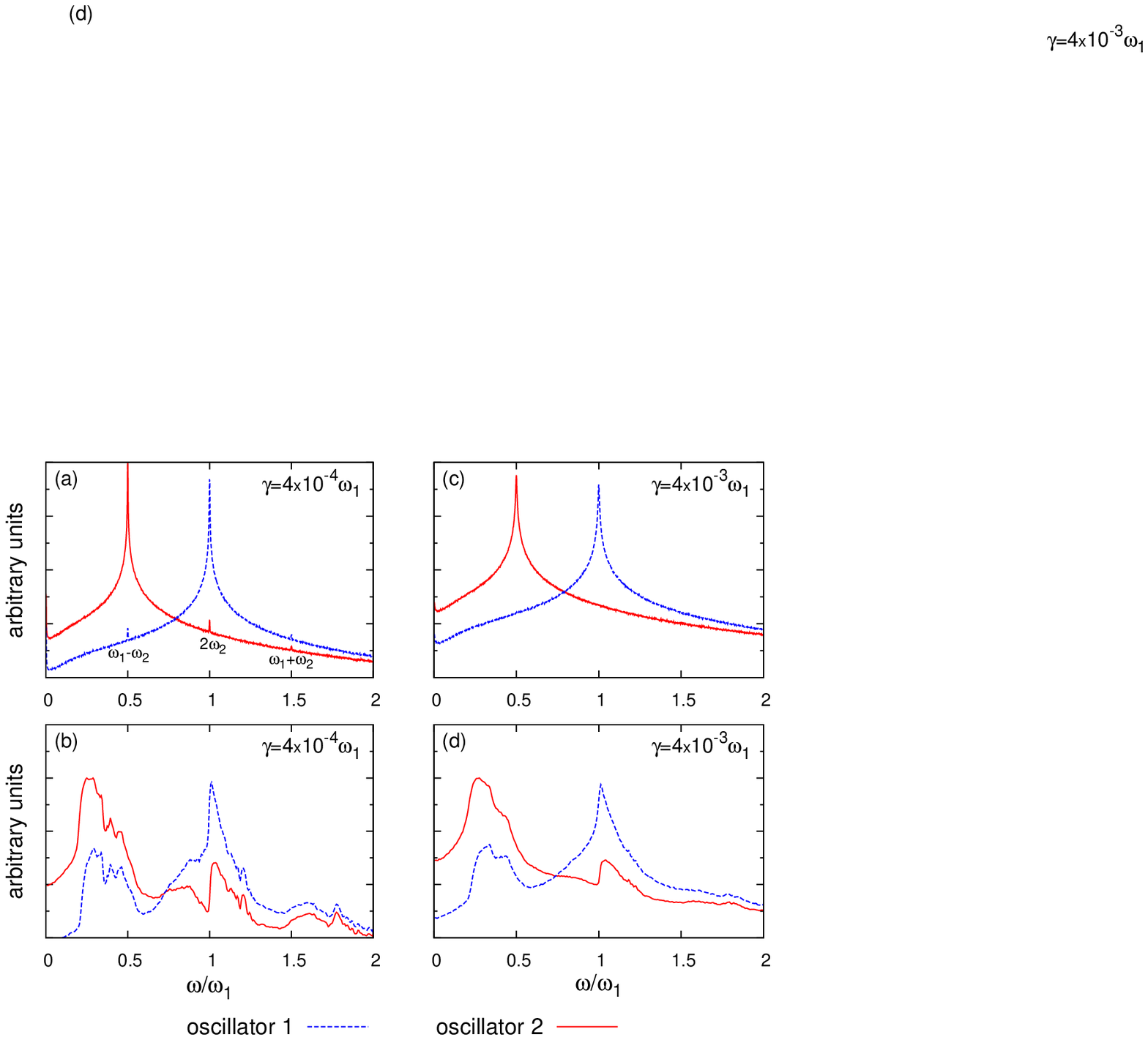}
\caption{Vibrational spectra (in logarithmic scale) of oscillators 1 and 2 obtained by QTB-MD simulation in the case of a cubic coupling (top panel) with $\Omega=0.5$ and $c_3=2.4\times 10^{-4}$, and in the case of a quartic coupling (lower panel) with $\Omega=0.2$ and $c_4=15.4\times 10^{-4}$. The spectra are computed for two selected values of the friction coefficient: $\gamma=4 \times 10^{-4} \omega_1$ and $\gamma = 4 \times 10^{-3}\omega_1$.}
\label{fig:sec2:spec}
\end{figure}
With this definition, the leakage is maximum when $\zeta=1$, i.e. when the system has reached an equipartition of the energy: $\epsilon_1^{\text{QTB}} = \epsilon_2^{\text{QTB}}$. In contrast, there is no leakage when $\zeta=0$, i.e. when $\epsilon_{1,2}^{\text{QTB}} =  \epsilon_{1,2}^{\text{exact}}$. 
In figure \ref{fig:sec2:zpel_vs_coupling}, the results obtained for $\zeta$ as a function of $\Omega$ for different values of $c_3$ and $c_4$ are presented.
One can note that the ZPEL strongly depends on the ratio of frequencies and is present only for certain values of $\Omega$. 
In the cubic case, it occurs only near $\Omega=0.5$ (figure \ref{fig:sec2:zpel_vs_coupling}.a). 
Indeed, cubic terms in the potential are known to be responsible for frequency doubling, that is the second harmonic generation ($2\omega$).  
This is confirmed by the vibrational spectrum of the two oscillators computed from QTB-MD in the cubic case (figure \ref{fig:sec2:spec}.a): harmonics at $2\omega_2$, $\omega_1-\omega_2$,  and $\omega_1+\omega_2$ are visible. Therefore, at $\Omega=0.5$, there is a resonance between the couple of modes $(\omega_1 ; 2\omega_2)$ and $(\omega_2 ; \omega_1-\omega_2)$.
Similarly, the quartic terms are responsible for the generation of modes with frequency $3\omega$; ZPEL is indeed observed near the resonance at $\Omega=1/3$ (figure \ref{fig:sec2:zpel_vs_coupling}.b).
With increasing quartic coupling, significant ZPEL also occurs for smaller values of $\Omega$.
Figure \ref{fig:sec2:spec}.b shows, in the case of $\Omega=0.2$ and $c_4=15.4\times 10^{-4}$, that many other modes than $\omega_1$ and $\omega_2$ also appear in the spectrum. Hence, multiple resonances are likely to occur leading to significant ZPEL for values of $\Omega < 1/3$.
%----------------------------
% FIGURE 4
%----------------------------
\begin{figure}[t]
\includegraphics[width=3.33in]{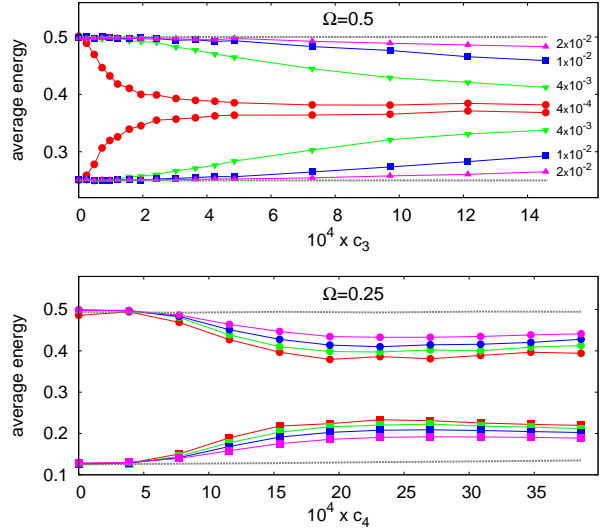}
\caption{Effect of the damping coefficient $\gamma$ (given in units of $\omega_1$) on the energies of the two oscillators (eq. \ref{energies}) as a function of the coupling constants. 
Top panel: cubic coupling ($c_3 \neq 0$, $c_4=0$) and $\Omega=0.5$. 
Lower panel: quartic coupling ($c_4 \neq 0$, $c_3=0$) and $\Omega=0.25$.  
The solid line and symbols represent the results obtained from QTB-MD and the grey dashed lines represent the exact results. \label{fig:sec2:ener_effet_gamma}}
\end{figure}

\paragraph{Influence of the friction coefficient $\gamma$}
The damping coefficient is now varied from $4 \times 10^{-4}\omega_1$ to $2 \times 10^{-2} \omega_1$. We focus on the frequency range where the ZPEL is important: $\Omega=0.5$ for cubic coupling and $\Omega=0.25$ for quartic coupling (see figure \ref{fig:sec2:zpel_vs_coupling}).
Figure \ref{fig:sec2:ener_effet_gamma} shows that the ZPEL strongly depends on $\gamma$. Increasing $\gamma$ can limit the leakage and even practically remove it in the case of the cubic coupling.  
In particular, for $c_3=2.4 \times 10^{-4}$, a value of $\gamma$ equal to $4 \times 10^{-3}\omega_1$ is sufficient to remove the ZPEL ($\zeta=0.08$). 
Figure \ref{fig:sec2:spec}.c shows the vibrational spectra obtained in this case with the larger $\gamma$: while the ZPEL has been suppressed, the peaks corresponding to the resonances ($2\omega_2$, $\omega_1-\omega_2$ and $\omega_1+\omega_2$) have disappeared. This further illustrates the relation between the mode resonances and the ZPEL. Moreover, increasing $\gamma$ also leads to a broadening of the peaks of the oscillators in the spectra, consistently with the fact that the full width at half maximum in the case of a harmonic oscillator is $\gamma/2\pi$ in a Langevin dynamics and for a spectrum in frequency.
The case of the quartic coupling is more complicated and even for large values of $\gamma$, the ZPEL is not completely suppressed (figure \ref{fig:sec2:ener_effet_gamma}). 
Figure \ref{fig:sec2:spec} also shows that increasing $\gamma$ in the case of a quartic coupling with $\Omega=0.2$ and $c_4=15.5 \times 10^{-4}$ is not sufficient to suppress all of the resonances between the different modes.

\begin{figure}
\centering
\includegraphics[width=3.33in]{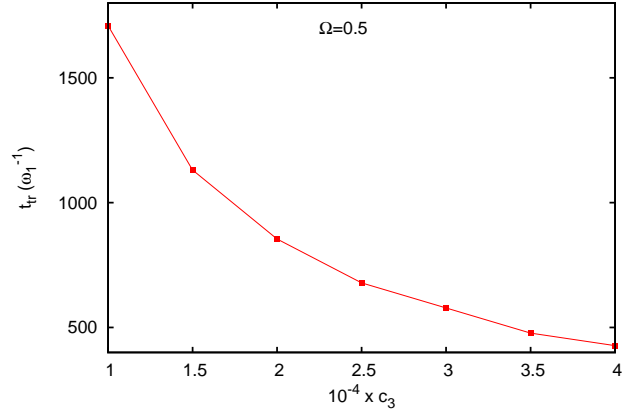}
\caption{Energy transfer time between the two oscillators $t_{tr}$ (in units of $\omega_1^{-1}$) as a function of the cubic coupling constant $c_3$ for $\Omega=0.5$ estimated from NVE simulations where only one oscillator is initially excited. Here, $c_4=0$.}
\label{fig:transfer}
\end{figure}

In order to estimate the characteristic time $t_{tr}$ of the energy transfer between the two oscillators, we performed NVE calculations where only oscillator 1 is initially excited. $t_{tr}$ can then be roughly estimated by calculating the typical time at which oscillator 2 starts to get excited.
Figure \ref{fig:transfer} shows the evolution of $t_{tr}$ for $\Omega=0.5$ as a function of the cubic coupling constant $c_3$. As expected, the characteristic time for transfer is directly related to the strength of the coupling.
To remove the ZPEL, we need to choose a value for $\gamma$ that is greater than the typical transfer frequency $\omega_\text{leakage}=1/t_{tr}$.
For example, in the case of $c_3=4\times 10^{-4}$, we find that $t_{tr}\sim 400\omega_1^{-1}$ and thus $\omega_\text{leakage}\sim 2.5\times 10^{-3}\omega_1$. 
Accordingly, figure \ref{fig:sec2:ener_effet_gamma} shows that a value of $\gamma=10^{-2}\omega_1$ or higher is necessary to remove the leakage i.e. the ZPEL is removed if $\gamma \gg \omega_\text{leakage}$.
%----------------------------
% FIGURE 5
%----------------------------

In conclusion, this simple model raises several important issues: the role of resonances and the possibility to remove or at least significantly reduce the effects of the ZPEL by increasing $\gamma$ beyond the typical frequencies for energy transfer between the modes. We now address these issues on a more complex model.

%************ Section3 : 1D chain  *****************
\subsection{One-dimensional chain of atoms}
We consider a one-dimensional chain of atoms, consisting of $3$ oxygen atoms interspaced with $3$ hydrogen atoms, with periodic boundary conditions. 
The interactions between the atoms are described by two interatomic potentials. 
On the one hand, the O--H interaction is a Morse-type potential derived by Johannsen for hydrogen-bonded systems:\cite{johannsen1998}
\begin{equation}
V_\text{OH}(r) = \frac{u_0}{a+b\mathrm{e}^{a(r-r_0)}} \left[ a\left( \mathrm{e}^{-b(r-r_0)}-1\right) + b \left( \mathrm{e}^{a(r-r_0)}-1\right) \right] - u_0
\label{eq:sec3:pot_oh}
\end{equation}
where $r$ is the O--H distance, $u_0$ is the height of the potential barrier, $r_0$ the equilibrium O--H distance, $a$ and $b$ are two parameters. 
The values of the parameters are set so that: $r_0 = 0.96$ \AA\, (which corresponds to the length of the covalent bond in the OH$^-$ ion), $a \simeq 7.11$ \AA$^{-1}$, $b \simeq 2.00$ \AA$^{-1}$ and $u_0 = 2.73$ eV so that the O--H stretching frequency ($\nu_\text{OH}$) in the harmonic approximation of the potential $V_\text{OH}$ approximately equals $100$ THz.
On the other hand, the O--O interaction is described by a standard Morse potential:
\begin{equation}
V_\text{OO}(R) = C_0\left(1-\mathrm{e}^{-\alpha_0 \left( R-R_0 \right) }\right)^2 - C_0
\label{eq:sec3:pot_oo}
\end{equation}
where $C_0$ and $\alpha_0$ are the depth and the width of the potential respectively and $R_0$ the O--O equilibrium distance. 
The parameters are the following: $C_0=3.81$ eV, $R_0=2.88$ \AA\, and $\alpha_0$ varies so that the value of the O--O frequency ($\nu_\text{OO}$) lies between $10$ and $60$ THz. The QTB-MD simulations are performed with a $0.1$ fs time step and equilibrium averaged values were obtained using 12 independent trajectories of $3$ ns each. 

The potential energy of an hydrogen atom is given by $ V_\text{OH}(r) + V_\text{OH}(R-r)$ which is a double-well potential. Within this model, we can define short "covalent" O--H bonds ($\sim 1$ \AA) and longer "hydrogen bonds" ($\sim 1.9$ \AA).
Although this model cannot represent a real physical system such as an ice cluster, it is characterized by realistic O-H frequencies and mode couplings and is useful to assess the nature and effects of ZPEL in realistic hydrogen-bonded systems.
A normal mode analysis of the system yields one low-frequency, $\nu_2$, corresponding to the O--O lattice mode, and two very similar optical high-frequencies, $\nu_1$, corresponding to the O--H stretching modes.
In analogy with the previous model, the O--H stretching modes roughly play the role of the high-frequency oscillator while the O--O lattice mode corresponds to the low frequency oscillator.
In the following, we show the influence of the parameter $\Omega= \nu_2/\nu_1$ and the friction coefficient $\gamma$ on the ZPEL at $T=600~$K. 
The frequency $\nu_2$ is varied through the parameter $\alpha_0$ (eq. \eqref{eq:sec3:pot_oo}) while $\nu_\text{OH}$ is fixed at $100$ THz (thus, the frequency $\nu_1$ is almost constant). 
The QTB-MD results are compared with those from PIMD simulations, using a Trotter number $P=20$ which ensures a good convergence of all the physical quantities in all cases studied here.
For each QTB-MD simulation, we checked that the total energy of the system, as well as the kinetic and potential energies, are in good agreement with the reference values given by PIMD.

\begin{figure}[t]
\centering
\includegraphics[width=3.33in]{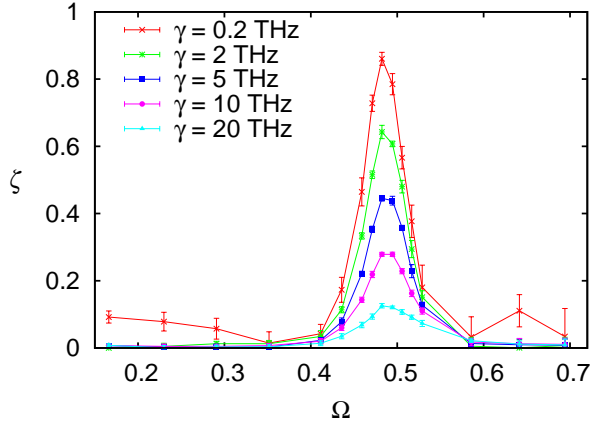}
\caption{ZPEL quantified by the deviation factor $\zeta$ using the definition in eq. \eqref{eq:sec3:zpel} as a function of the frequency ratio $\Omega$.}
\label{fig:sec3:zpel}
\end{figure}

In order to evaluate the leakage, we compare the kinetic energy of the light atoms, significantly involved in the high-frequency modes, to that of the heavier atoms, mainly involved in the low-frequency modes. Thus, the effective temperatures $T_\text{H}$ and $T_\text{O}$ of H and O atoms are defined from the kinetic energies: 
\begin{equation}
 \frac{k_B T_\text{H}}{2} = \frac{1}{N_\text{H}} \sum_{i=1}^{N_\text{H}} \langle E_k^{(i)} \rangle, \quad \frac{k_B T_\text{O}}{2} = \frac{1}{N_\text{O}} \sum_{i=1}^{N_\text{O}}  \langle E_k^{(i)} \rangle
\end{equation}
where $N_\text{H}=3$ and $N_\text{O}=3$ are the numbers of H and O atoms respectively, and $\langle E_k^{(i)} \rangle $ the average kinetic energy of atom $i$. 
%
%-----------------------
% FIGURE 6
%-----------------------
\begin{figure}[t]
\centering
\includegraphics[width=3.33in]{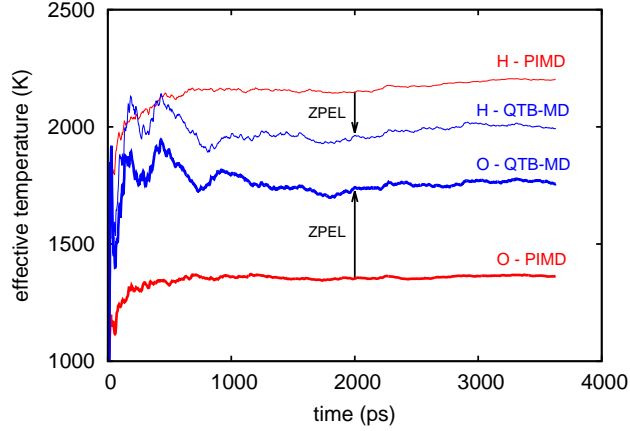}
\caption{Effective temperatures $T_\text{O}$ and $T_\text{H}$ of oxygen and hydrogen atoms calculated with QTB-MD and PIMD at $T=600K$. Here $\Omega=0.5$ and $\gamma=0.2$ THz. The arrows indicate the temperature shifts due to the ZPEL.}
\label{fig:sec3:temp}
\end{figure}
In a classical system, equipartition ensures that the kinetic energy is equally distributed among all degrees of freedom: they all have the same effective temperature.
This is not true in the quantum case: high-frequency modes have more kinetic energy and their effective temperature is therefore greater. This is the case for the QTB method and for PIMD, which serves as a reference here. 
From figure \ref{fig:sec3:temp}, one sees that, as expected, the leakage tends to increase the effective temperature of light atoms and decrease that of heavier atoms. 
In this case, the ZPEL can be quantified through the deviation factor:
\begin{equation}
\zeta = \frac{\left( T_\text{H}-T_\text{O}\right)^{(\text{PIMD})} - \left( T_\text{H}-T_\text{O}\right) ^{(\text{QTB})} }{\left( T_\text{H}-T_\text{O}\right) ^{(\text{PIMD})}}.
\label{eq:sec3:zpel}
\end{equation}
$\zeta=0$ if there is no leakage and $0<\zeta<1$ if leakage occurs and its dependence on $\Omega$ is shown in figure \ref{fig:sec3:zpel}.
Similarly to the coupled harmonic oscillators' model (section \ref{sec_2}), ZPEL occurs mostly for $\Omega\sim 1/2$.
Figure \ref{fig:sec3:zpel} also shows that the ZPEL can be substantially decreased by increasing $\gamma$, as in the previous model. 
On the other hand, important ZPEL is observed for $\Omega< 0.2$: this corresponds to a highly anharmonic regime where a structural transition occurs and therefore corresponds to a different physical situation than the other values of $\Omega$. 
%----------------------
% FIGURE 7
%----------------------

\paragraph{ZPEL effects on structural properties}
%-------------------
% FIGURE 8
%-------------------
\begin{figure}[t]
\centering
\includegraphics[width=3in]{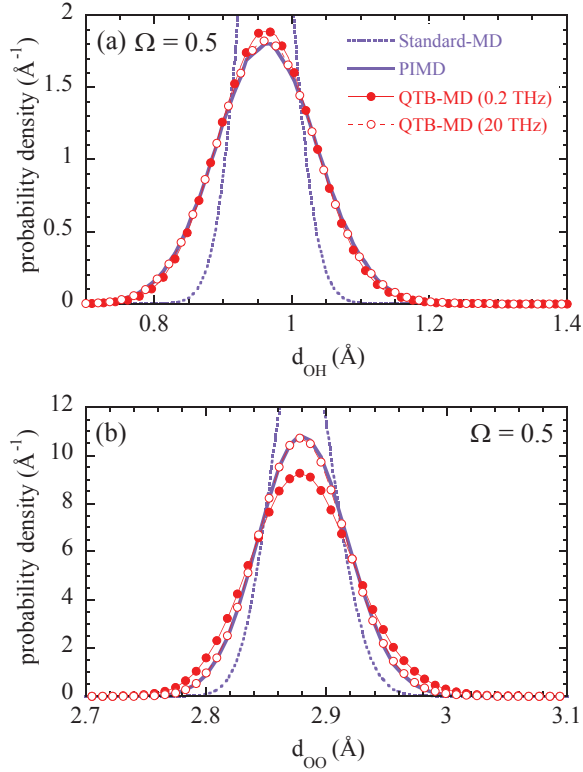}
\caption{Probability distribution of the O--H (a) and the O--O (b) distances computed from standard-MD, PIMD and QTB-MD, for selected values of the friction coefficient $\gamma$ ($0.2$ and $20$ THz) and for $\Omega=0.5$.}
\label{fig:sec3:pcf}
\end{figure}
Figure \ref{fig:sec3:pcf} shows the distributions of interatomic distances, $d_\text{OH}$ and $d_\text{OO}$ for the case $\Omega=0.5$ computed from QTB-MD, PIMD and standard-MD simulations.
In figure \ref{fig:sec3:pcf}.a, one can see that the $d_\text{OH}$ distribution is almost not affected by the ZPEL. 
On the other hand, the $d_\text{OO}$ distribution is more sensitive to the ZPEL: the QTB-MD distribution is too broad, which is consistent with the excess of kinetic energy for the oxygen atoms that comes from the ZPEL. However, when the ZPEL is suppressed, by increasing $\gamma$, the QTB-MD $d_\text{OO}$ distribution coincides with the PIMD one.

\begin{figure}
\centering
\includegraphics[width=3.33in]{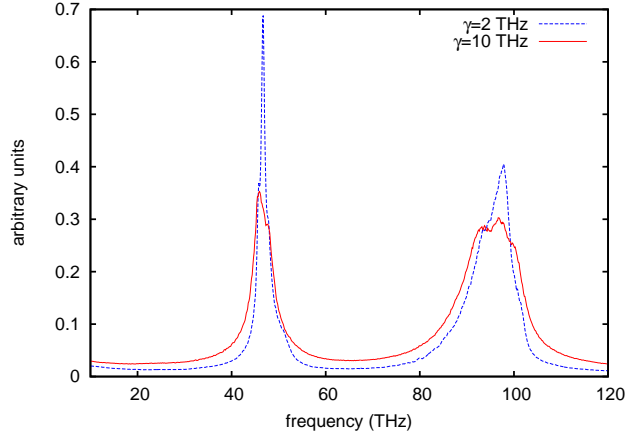}
\caption{Vibrational spectra obtained by QTB-MD simulation for $\Omega=0.5$ and two values of the friction coefficient: $\gamma=0.2$ THz (blue dashed line) and $\gamma=10$ THz (red full line).}
\label{fig:sec3:spec}
\end{figure}

\paragraph{ZPEL effects on vibrational properties}
We have seen in the case of the two coupled harmonic oscillators that increasing $\gamma$ has consequences on the vibrational spectrum of the system; in particular, the peaks are broadened and the peaks corresponding to the mode resonances disappear when $\gamma$ is large enough (see figure \ref{fig:sec2:spec}).
Figure \ref{fig:sec3:spec} shows the vibrational spectrum of the one-dimensional chain of atoms for $\Omega=0.5$ and for two different values of $\gamma$ ($0.2$ and $10$ THz). For $\gamma=0.2$ THz, ZPEL occurs while for $\gamma=10$ THz, the ZPEL is almost fully removed (see figure \ref{fig:sec3:zpel}). We can see that increasing the friction coefficient leads to broader peaks as expected. However, the positions of these peaks hence the mode frequencies, are not modified by the large value of $\gamma$. 
Even with a large damping term, the vibrational spectrum still yields useful information about the mode frequencies in this case. 
%-------------------
% FIGURE 9
%-------------------

In conclusion, in a system containing different chemical elements, the kinetic energy ratio between them can be used as an indicator of the ZPEL.
As in the case of two coupled harmonic oscillators, the ZPEL is intrinsically related to resonances between vibrational modes. Correlatively, increasing the friction coefficient allows to remove the ZPEL. 
In this case, the quantum structural properties are well reproduced; the dynamical properties should be regarded with caution but the vibrational spectrum still contains useful information.

%************ Section4 : Aluminium+BTO  *****************
\section{Applications to realistic systems} \label{sec_3}
We now investigate the effect of the friction coefficient $\gamma$ on the ZPEL for systems with many degrees of freedom.

\subsection{Lennard-Jones Aluminium}
\begin{figure}
\centering
\includegraphics[width=3in]{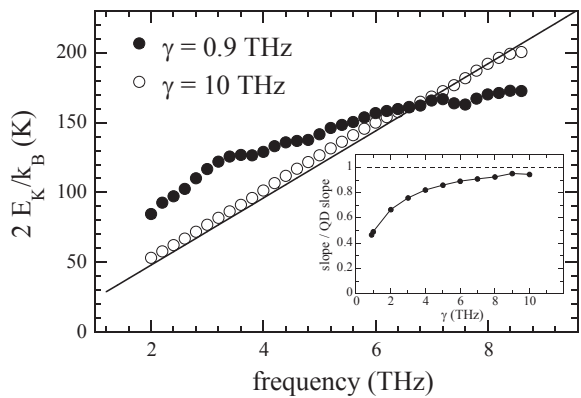}
\caption{Scaled kinetic energy distribution of aluminum at $T=10~$K as a function of the frequency of the modes, obtained from QTB-MD simulation with $\gamma$ values of 0.9 and 10~THz, while the frequency cut-off ($20~$THz) is chosen equal to twice the highest frequency of the system. The solid line corresponds to the quantum distribution (QD) $\theta(\nu,T)$ (eq. \ref{eq-theta} with $\nu=\omega/2\pi$). Inset: evolution as a function of $\gamma$ of the slope of the energy distribution, normalized according to the quantum distribution.}
\label{fig:sec4:al}
\end{figure}

Using QTB-MD simulations, Bedoya-Mart\'inez \textit{et al.}\cite{bedoya2014} have evidenced the ZPEL at $T=10$ K in a crystal of aluminium modeled by a Lennard-Jones potential ($\epsilon/k_B = 1450.6$ K, $\sigma =2.54$ \AA, $\text{cutoff}=1.37\sigma=3.49$ \AA). In their paper, they showed that the energy is transferred from the high-energy modes to the low-energy modes because the QTB method is unable to fully counterbalance this leakage. 
We carried out QTB-MD simulations using a $1$ fs time step and different values of $\gamma$.  
We confirm that, with $\gamma=0.9$ THz, QTB-MD fails to give the correct quantum energy distribution, as illustrated by the full circles in figure \ref{fig:sec4:al}. Indeed, the resulting distribution is intermediate between the quantum and the classical homogeneous distributions.
However, with a higher value of $\gamma$ (10 THz), the energy distribution from QTB-MD is very close to the expected quantum distribution $\theta(\nu,T)$ (eq. \ref{eq-theta}), as given by the open circles in figure \ref{fig:sec4:al}. 
Therefore, for large enough damping, the ZPEL is neutralized by the QTB. The inset of figure \ref{fig:sec4:al} provides the evolution of the slope of the energy distribution, normalized by that of the quantum distribution, as a function of $\gamma$. The larger the friction coefficient, the lower the ZPEL, up to $\gamma=9$ THz for which a plateau value is reached. From this value of $\gamma$ upwards, the leakage is satisfactorily reduced and the energy distribution obtained by QTB-MD is the one initially introduced in the colored noise.
%-----------------------
% Figure 10
%-----------------------

\begin{figure}
\centering
\includegraphics[width=3in]{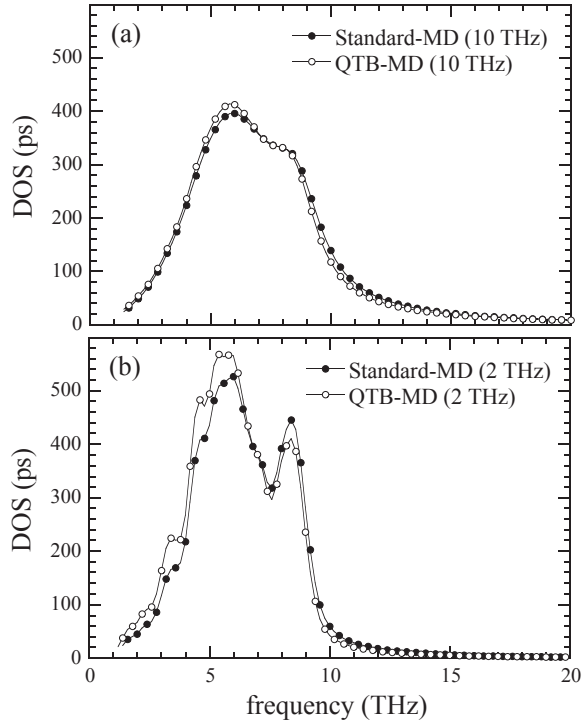}
\caption{Vibrational density of states (DOS) as a function of the frequency for different values of $\gamma$. They have been computed through the Fourier transform of the velocity autocorrelation function which is normalized by $k_BT$ in the case of Langevin MD or by $\theta(\nu,T)$ (eq. \ref{eq-theta}) in the case of QTB-MD. Two values of the friction coefficient $\gamma$ are used: (a) $10~$THz and (b) $2~$THz.}
\label{fig:sec4:dos}
\end{figure}

The disadvantage of any thermostat involving a damping term, as in the QTB or the Langevin thermostat, is the possible broadening of the vibrational peaks and the possible occurrence of a spurious high-frequency tail in the phonon density of states (DOS).  
For small values of $\gamma$, i.e. when $\gamma$ is lower than the full width at half maximum ($\Delta\omega$) of the peaks of the DOS, increasing $\gamma$ does not significantly perturb the spectrum. Conversely, for large values of $\gamma$, the broadening induced by the damping term is of the order of $\Delta\nu=\Delta\omega/2\pi=\gamma/2\pi$.
This issue is shown in figure \ref{fig:sec4:dos}. In the case of standard MD with a Langevin thermostat, the DOS is obtained by normalizing the Fourier transform of the velocity autocorrelation function by $k_BT$. In the case of QTB-MD, $k_BT$ must be replaced by $\theta(\omega,T)$ (eq. \ref{eq-theta}). Figure \ref{fig:sec4:dos}.a) shows that, when the ZPEL is removed, the DOS obtained from QTB-MD trajectories is close to that derived from standard-MD.
In contrast, figure \ref{fig:sec4:dos}.b) shows that, when ZPEL occurs, the DOS cannot be obtained from the QTB-MD trajectories, since the number of high-frequency modes or low-frequency modes are underestimated and overestimated, respectively.
%---------------------
% Figure 11
%---------------------

\subsection{Barium titanate}
\label{sec_bto}
BaTiO$_3$ (BTO) is a strongly anharmonic ferroelectric crystal characterized by a complex energy landscape. 
Moreover, quantum effects have been shown to influence its structural properties. \cite{gen,volk}
It undergoes a complex sequence of structural phase-transitions \cite{ferro} as temperature increases: from rhombohedral (R), to orthorhombic (O), tetragonal (T), and cubic (C) structures. 
Each of these phase transitions coincides with the temperature at which the local modes (dipoles) move out of the potential wells in which they were confined, and visit a new potential energy minimum, giving rise to a new value and direction of the macroscopic polarization. 
Such a behavior is a challenge for the QTB approach because of the intrinsic anharmonicity of the system.

QTB-MD simulations were performed for temperatures ranging from 1 K to 270 K, using a Langevin barostat \cite{gen} {whose equations are given in Appendix \ref{appendixB}}. The ferroelectric properties of BTO were modeled by an effective {Hamiltonian \cite{sto1,sto2}} derived from first-principles density-functional calculations. The degrees of freedom of this Hamiltonian are the local modes and the (homogeneous) strain tensor. 
The friction coefficient $\gamma$ was varied from 0.5 to 16 THz, while the cut-off frequency $\nu_\text{cut}$ is chosen equal to four times the maximum frequency in the system (5~THz). Here, we investigate the convergence, with respect to $\gamma$, of the values of the three phase-transition temperatures (R-O, O-T, T-C).
%-----------------
% Figure 12
%----------------
\begin{figure}
\centering
\includegraphics[width=3in]{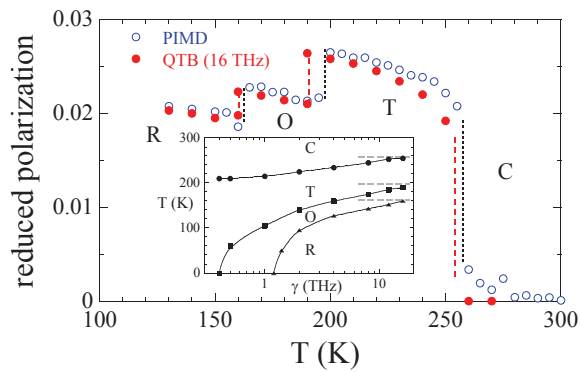}
\caption{Temperature evolution of the reduced polarization associated with the ferroelectric transition in BaTiO$_3$, as obtained by QTB-MD ($\gamma=16$~THz) and PIMD ($P=16$) simulations. Vertical dashed lines show the transition temperatures obtained for the R-O, O-T, and T-C transitions. The inset provides the convergence of the transition temperatures with the frictional coefficient, $\gamma$, of the QTB method. The horizontal grey dashed lines give the temperatures obtained by PIMD.}
\label{fig:sec4:bto}
\end{figure}
Figure \ref{fig:sec4:bto} displays the reduced polarization (see reference \cite{gen}) as a function of the temperature obtained by QTB-MD with $\gamma=16$~THz. For this damping value, the QTB-MD simulation (full circles) gives the expected sequence of phase transitions: R-O-T-C, in agreement with the converged PIMD result with a Trotter number $P=16$ (open circles). The three consecutive transition temperatures: 160~K, 190~K, and 255~K are similar to those obtained by PIMD (163~K, 198~K, and 258~K, respectively). The inset in figure \ref{fig:sec4:bto} shows the convergence of the transition temperatures as a function of $\gamma$ within QTB-MD. For low $\gamma$ values, the rhombohedral and orthorhombic phases are missed. It is worth noting that for large $\gamma$ values, the QTB method yields the correct series of phase transitions: the effects of the ZPEL have been suppressed.

%************ Conclusion  *****************
\section{Conclusion and practical consequences} \label{sec_concl}
We have performed a systematic and quantitative study of zero-point energy leakage (ZPEL) in QTB-MD simulations. The aim here is to assess the reliability of the QTB method on various systems with different degrees of complexity. We have found that the ZPEL is intrinsically related to resonances between vibrational modes and, as in realistic systems many modes can resonate, this is hardly avoidable in practice. However, increasing the damping term $\gamma$ significantly reduces the leakage and can even, in some cases, remove it entirely. A striking example is provided by our results on BTO, as with small damping term, the phase diagram obtained by the QTB method is wrong, while for larger damping, one recovers the complete sequence of phase transitions at the correct temperatures.

This effect can be explained as follows. The QTB method connects a classical system to a thermal bath which follows the quantum fluctuation-dissipation theorem. Therefore, there is no equipartition of the energy since the QTB pumps more energy into high-frequency modes than low-frequency ones. The ZPEL results from the transfer of energy from high-frequency to low-frequency modes: the obtained energy distribution is therefore the result of the balance between QTB pumping and damping on the one hand and energy transfer within the system on the other hand. Increasing the damping term will increase the pumping rate with respect to the internal equilibration and the QTB energy distribution becomes closer to the quantum one. Moreover, when $\gamma$ is larger than the characteristic frequency of the energy transfer between vibrational modes, the effects of the resonances between these modes are hindered. Hence, a simple and effective way to prevent ZPEL to occur in QTB-MD simulations is to increase the damping term $\gamma$.

This raises the issue that within the frame of a Langevin simulation, one should decrease, not increase, the damping term in order not to alter the dynamics of the system too dramatically. Careful analysis of the effect of damping on both structural and dynamical properties tends to show that this question should be addressed with care in each specific case, but that QTB-MD simulations turn out to be relatively robust and yield excellent results as long as one keeps in mind the physics of the problem. Indeed, we have seen that the mode frequencies obtained from QTB-MD vibrational spectra are not dramatically altered by the increase of $\gamma$, even though a large $\gamma$ implies a broadening of the peaks. This allows us, for example, to study the O-H stretching or bending modes in hydrogen-bonded materials since the corresponding frequencies are usually much larger than $\gamma$. On the contrary, we expect the low frequency part of the spectrum to be substantially affected by a large damping term.

Therefore, the QTB method is an efficient tool to study a large variety of anharmonic systems provided that the value of the friction coefficient is large enough to ensure that the ZPEL remains negligible. In this case, the QTB method presents several advantages compared to path integral methods : its computational cost is similar to that of standard MD simulations, enabling the study of large and complex systems, and dynamical properties are directly accessible making possible the confrontation of QTB-MD results to spectroscopic measurements for example.

\begin{acknowledgement}
We gratefully acknowledge important contributions to this work by Grégory Geneste, who provided us the code to perform the calculations on BTO and enlightened many issues about ferroelectric transitions. We also thank Jean-Louis Barrat for providing the values of the LJ parameters they used in their paper to describe the Aluminium crystal. Y.B. acknowledges financial support from the Conseil Régional d'Île-de-France through the DIM Oxymore. This work was performed using HPC resources from the "mésocentre" computing center of CentraleSupélec, which is supported by CentraleSupélec and CNRS.
\end{acknowledgement}

%************ Appendix A  *****************
\appendix
\section{Generation of the random force}\label{appendixA}
{
This part presents the technique used to generate the fluctuating force R(t) which is a random gaussian variable.
This technique has been proposed by Dammak \textit{et al.} \cite{dammak2009} and is based on a procedure proposed by Maradudin \textit{et al.} to generate random surfaces of specific roughness\cite{maradudin}. 
Here, we want to generate the stationary Gaussian process $R(t)$ with the following properties:
\begin{align}
\langle R(t) \rangle &= 0\\
\langle R(t)R(t+\tau) \rangle &= \int_{-\infty}^{+\infty}I_R(\omega,T)\text{e}^{-i\omega \tau}\frac{\text{d}\omega}{2\pi} \label{eq-autocorrel}
\end{align}
The second equation is the Wiener-Khinchin theorem that relates the autocorrelation of $R(t)$ to its power spectral density, $I_R(\omega,T)$. In QTB-MD simulations, $I_R$ is given by eq. \ref{eq-IR}. 
The value of the random noise at a time $t_n=n\delta t$ is a Gaussian random variable that can be written as a sum of independent Gaussian random variables $X_j$ with weights $W_j$ as 
\begin{equation}
R_n \equiv R(t_n)=\sum_{j=-\infty}^{+\infty} W_j X_{j+n} \label{eq-R}
\end{equation}
The variables $X_j$ have zero mean and a standard deviation of unity.
From eqs. \ref{eq-autocorrel} and \ref{eq-R}, the weights $W_j$ obey the following relation:
\begin{equation}
\sum_{j=-\infty}^{+\infty} W_j W_{j-l} =\int_{-\infty}^{+\infty}I_R(\omega,T)\text{e}^{-i\omega t_l}\frac{\text{d}\omega}{2\pi} \label{eq-WW}
\end{equation}
with $t_l=l\delta t$. $\tilde{W}(\omega)$ is defined as the Fourier transform of $W(t)$:  
\begin{equation}
W_j \equiv W(t_j)=\int_{-\infty}^{+\infty}\tilde{W}(\omega)\text{e}^{-i\omega t_j}\frac{\text{d}\omega}{2\pi} \label{eq-W-1}
\end{equation}
Using expression (\ref{eq-W-1}), in the continuous limit ($\delta t \rightarrow 0$):
\begin{equation}
\sum_{j=-\infty}^{+\infty} W_j W_{j-l}=\frac{1}{\delta t}\int_{-\infty}^{\infty}\tilde{W}(\omega)\tilde{W}(-\omega)\text{e}^{-i\omega t_l}\frac{\text{d}\omega}{2\pi} \label{eq-WW-2}
\end{equation}
Considering $W(t)$ as even and real, the function $\tilde{W}(\omega)$, also even and real, is obtained using eq. \ref{eq-WW} and \ref{eq-WW-2}:
\begin{equation}
\tilde{W}(\omega)=\sqrt{\delta t I_R(\omega,T)}
\end{equation}
and:
\begin{equation}
W_j=\sqrt{\delta t}\int_{-\infty}^{+\infty}\sqrt{I_R(\omega,T)}\text{e}^{-i\omega t_j}\frac{\text{d}\omega}{2\pi} \label{eq-W-2}
\end{equation}
In practice, the time and  the pulsations are discretized. The Fourier transform are expressed so that
\begin{align}
W_j&=\frac{1}{\sqrt{N}}\sum_{l=-N/2+1}^{N/2}\tilde{W}_l \text{e}^{-i2\pi jl/N} \label{eq-W-DFT}\\
X_j&=\frac{1}{\sqrt{N}}\sum_{l=-N/2+1}^{N/2}\tilde{X}_l \text{e}^{-i2\pi jl/N} \label{eq-X-DFT}
\end{align} 
with $N$ the total number of MD steps. By comparing eq. \ref{eq-W-2} with the integral form of eq. \ref{eq-W-DFT}:
\begin{equation}
W_j=\frac{1}{\sqrt{N}\delta\omega}\int_{-\infty}^{+\infty}\tilde{W}(\omega)\text{e}^{-i\omega t_j}\text{d}\omega
\end{equation}
and using $N\delta t\delta \omega=2\pi$, we obtain that
\begin{equation}
\tilde{W}_l=\frac{1}{\sqrt{N\delta t}}\sqrt{I_R(\omega_l)} \label{eq-W-3}
\end{equation}
with $\omega_l=l\delta \omega$. From eq. \ref{eq-W-DFT} and \ref{eq-X-DFT}, $R_n$ writes
\begin{equation}
R_n=\sum_{l=-N/2+1}^{N/2} \tilde{W}_{-l}\tilde{X}_l\text{e}^{-i2\pi nl/N}
\end{equation}
and using eq. \ref{eq-W-3}, $R_n$ finally is:
\begin{equation}
R_n=\frac{1}{\sqrt{N\delta t}}\sum_{l=-N/2+1}^{N/2}\sqrt{I_R(\omega_l)}\tilde{X}_l\text{e}^{-i2\pi nl/N}
\end{equation}
Let us define the discrete Fourier transform $\tilde{R}_l$ such that 
\begin{equation}
R_n=\frac{1}{\sqrt{N}}\sum_{l=-N/2+1}^{N/2}\tilde{R}_l\text{e}^{-i2\pi nl/N} \label{eq-Rn}
\end{equation}
and then obtain that 
\begin{equation}
\tilde{R}_l=\sqrt{\frac{I_R(\omega_l)}{\delta t}}\tilde{X}_l
\end{equation}
the Gaussian random variables $\tilde{X}_l$ can be rewritten as
\begin{equation}
\tilde{X}_l=\frac{\tilde{M}_l+i\tilde{N}_l}{\sqrt{2}}
\end{equation}
with $\tilde{M}_l$ and $\tilde{N}_l$ independent Gaussian random variables with zero mean and a standard deviation of unity. 
Moreover, to ensure that the variables $\tilde{X}_l$ are real, $\tilde{M}_l=\tilde{M}_{-l}$ and  $\tilde{N}_l=-\tilde{N}_{-l}$ are required.
Finally:
\begin{equation}
\tilde{R}_l=\sqrt{\frac{I_R(\omega_l)}{2\delta t}}\left(\tilde{M}_l+i\tilde{N}_l\right) \label{eq-R-tilde}
\end{equation}
In practice, the random forces $R_n$ are obtained using the following steps:
\begin{enumerate}
\item Generation of independent Gaussian random numbers $\tilde{M}_l$ and $\tilde{N}_l$ for $l=1,..,N/2-1$
\item Computation of $\tilde{R}_l$ using expression (\ref{eq-R-tilde})
\item Symmetrization of $\tilde{R}_l$ : $\tilde{R}_l=\tilde{R}_{N-l}$ for $l=N/2+1,..,N-1$
\item Cancellation ($\tilde{R}_l=0$) for $l=0$ and $N/2$
\item Computation of $R_n$ from eq. \ref{eq-Rn}
\end{enumerate}
                                                                                 }

%************ Appendix B  *****************
\section{Langevin barostat}\label{appendixB}
{
This section gives the equations of the Langevin barostat used to fix the hydrostatic pressure in BTO.
The extension of the Langevin method to the isothermal-isobaric ensemble has been achieved by Quigley and Probert\cite{qp1,qp2}, 
giving rise to an algorithm in which random and friction forces are applied, not only on the atomic coordinates, 
but also on the supercell vectors. 
In the following expressions, second-rank tensors are written in bold. 
The equations of motion on the local mode $i$ (with mass $m$) using the Langevin barostat are:
\begin{equation}
\frac{ d \vec p_i} { dt} = \vec f_i - \gamma  \vec p_i  + \vec R_i - \frac{\bf p_G}{W_g} \vec p_i - \frac{1}{N_f} . \frac{Tr({\bf p_G})}{W_g} \vec p_i
\end{equation}
with $\vec f_i = - \vec \nabla_{\vec u_i} \Phi(\vec u_1,...,\vec u_N)$ the internal force. 
The terms $- \gamma \vec p_i$ and $\vec R_i$ correspond to the friction and the random forces of the thermostat (Langevin or QTB).
The momentum $\vec p_i$ is related to the position $\vec u_i$ by
\begin{equation}
\frac{d \vec u_i}{dt} = \frac{\vec p_i}{m} + \frac{\bf p_G}{W_g} \vec u_i
\end{equation}
while the matrix of the supercell vectors ${\bf h}$ and its conjugate momentum ${\bf p_G}$ evolve according to
\begin{equation}
\frac{d{\bf h}}{dt} = \frac{\bf p_G h}{W_g}
\end{equation}
and
\begin{equation}
\label{eq1}
\frac{d {\bf p_G}}{dt} = V(t)({\bf X} - P_{ext} {\bf Id}) + \frac{1}{N_f} \sum_{i} \frac{ {\vec p_i^{2}} } {m}  {\bf Id} - \gamma_G  {\bf p_G} + {\bf L_G}
\end{equation}
in which $V(t)$ is the supercell volume, $W_g$ is the "mass" associated to the barostat, $N_f$ is the number of degrees of freedom, $P_{ext}$ is the external pressure, ${\bf Id}$ is the identity tensor and ${\bf X}$ is the internal pressure tensor\cite{qp1}.
In the right member of Eq.~\ref{eq1}, one recognizes a friction force on the supercell $- \gamma_G  {\bf p_G}$ ($\gamma_G$ is a friction coefficient for the barostat) and a random force ${\bf L_G}$, a 3 $\times$ 3 matrix whose components are randomly drawn at each time step in a gaussian with variance  $\sqrt{\frac{2 \gamma_G W_g k_B T}{\delta t}}$. This random force on the barostat is symmetrized at each time step to avoid global rotation of the supercell during the simulation.                                                                                 }

%************ Bibliography  *****************

% ----------- Figure for TOC --------------
\begin{tocentry}
\begin{center}
\includegraphics[scale=0.267]{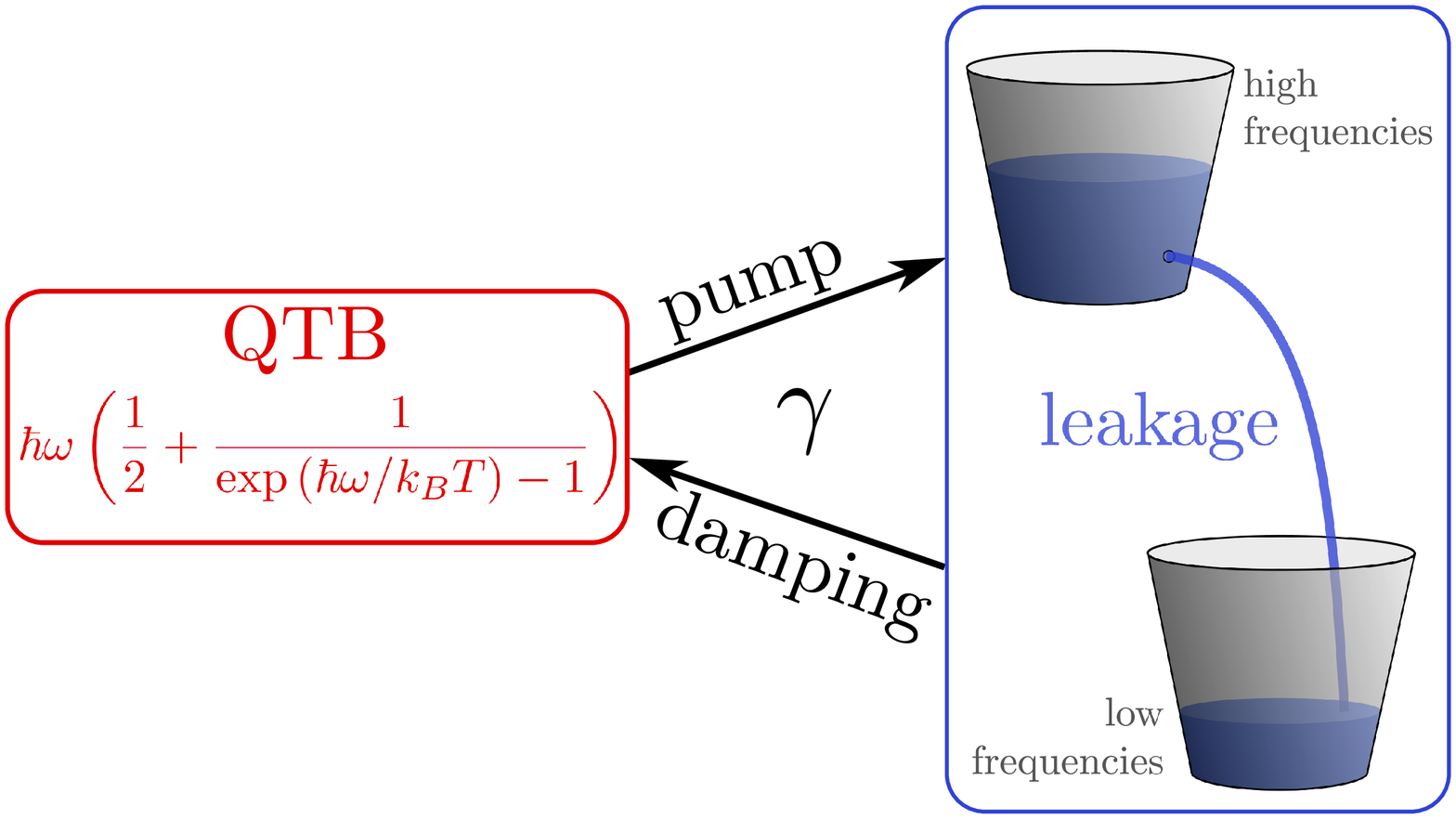}
\end{center}
\end{tocentry}

\end{document}